\documentclass[preprint,showpacs]{revtex4}
\usepackage{graphicx}%
\usepackage{dcolumn}
\usepackage{amsmath}
\usepackage{latexsym}

\begin{document}
\title{Energy, Entropy and the Ricci Flow}  
\author{Joseph Samuel and Sutirtha Roy Chowdhury}
\affiliation{Raman Research Institute, Bangalore-560 080}
\date{\today}
\begin{abstract}

The Ricci flow is a heat equation for metrics, which
has recently been used to study the topology of closed three
manifolds.
In this paper we apply Ricci flow techniques to general relativity.
We view a three dimensional asymptotically flat Riemannian 
metric as a time symmetric initial data set for Einstein's equations. 
We study the evolution of the area ${\cal A}$ and Hawking mass ${\cal 
M}_H$ of a two dimensional closed surface under the Ricci flow. 
The physical relevance of our study derives
from the fact that, in general relativity the area of apparent horizons 
is related to black hole entropy and the Hawking mass of an asymptotic 
round 2-sphere is the ADM energy. 
We begin by considering the special case of 
spherical symmetry to develop a physical feel for the geometric
quantities involved. We then consider a general    
asymptotically flat Riemannian metric
and derive an inequality
$\frac{d}{d\tau}{\cal A}^{3/2}\le -24\pi^{3/2} {\cal M}_H$ which relates 
the evolution of the area of a closed surface $S$ to its Hawking mass.
We suggest that there may be a maximum principle which governs
the long term existence of the asymptotically flat Ricci flow.

\end{abstract}

\maketitle

\section{Introduction}
The Ricci flow \cite{friedan,hamilton,perelman,cao} 
has been used by mathematicians to understand the topology of
three manifolds. It appears likely that these mathematical
developments will also be useful in physics in the study of
geometric theories like general relativity. Several papers 
\cite{eric1,eric2,eric3,eric4,eric5,wiseman,soludukhin} have appeared
dealing with physical applications of the Ricci flow. 
The Ricci flow (RF) is a (degenerate) parabolic differential equation, 
and is very similar to the heat equation. In a previous paper 
\cite{black} we explored an analogy between the 
Ricci flow and thermodynamics. The analogy is based on the 
observation that the Ricci flow (like the heat equation)
loses memory of initial conditions, just as a physical system
loses memory of its initial state as it approaches thermal equilibrium.
As was noticed there, a slight
modification of the Ricci flow yields Schwarzschild
space as  a fixed point. 
In this paper we look at the unmodified Ricci flow to
see how some physically interesting quantities evolve with the flow. 
Energy and entropy are quantities of physical interest from the
thermodynamic point of view.
In general relativity these quantities take on a purely differential 
geometric meaning: the entropy is related to the area of black 
hole horizons
and the energy to the ADM mass at infinity. We investigate the 
the evolution of these quantities under the Ricci flow 
and derive some inequalities relating them. 

The subject of entropy bounds \cite{bekenstein,bousso} 
has been of wide and sometimes controversial \cite{unruh} interest to 
physicists. It is believed that such bounds 
contain clues towards quantum gravity. Entropy bounds 
motivate ideas like ``the holographic hypothesis''. 
There is a geometric entropy bound proposed by Penrose, which connects 
two geometrical quantities: the area of the outermost horizon and the ADM 
mass. 
In the seventies, Roger Penrose\cite{penrose}, in an attempt to 
test the idea of cosmic censorship (which is part of the
``establishment view'' of gravitational collapse), 
used physical arguments to deduce from cosmic censorship
an inequality relating the ADM mass of an initial data set for GR and 
the area of its outermost apparent horizon: $M_{ADM}\ge 
\sqrt{16\pi {\cal A}}$. 
This inequality, which is saturated by the Schwarzschild space,
has a clear thermodynamic interpretation. 
It states that Schwarzschild spacetime maximises geometric
entropy for a given energy. Maximising entropy for fixed energy 
is a property which characterises a thermal state in statistical 
mechanics. The thermal character of Hawking 
radiation from a Schwarzschild black hole is entirely consistent with 
this interpretation.
Penrose's inequality appears to capture something
deep about general relativity, with a thermodynamic statement 
expressed in geometric terms. In this respect, it is similar to the
the area theorem\cite{wald},
which represents the second law of thermodynamics in geometric form. 

A counterexample to Penrose's inequality would imply a flaw in the 
establishment view. No counterexample has so far been found. Special
cases of Penrose's inequality have been proved\cite{bray,bray2,huisken} using geometric flow techniques. 
Jang and Wald 
\cite{jang} showed that 
if one assumes the existence of the inverse mean curvature (IMC) 
flow, (a particular diffeomorphism of a spatial slice), 
one could address Penrose's
inequality in the special case of time symmetric initial data. 
Their work was based on the monotonicity of the Hawking mass under
the inverse mean curvature flow, which follows from earlier work by 
Geroch\cite{geroch}. 
This suggests a proof of the Penrose inequality \cite{jang}.
The only gap in this proof is the existence of the IMC flow.
This gap was filled by Huisken and Ilmanen \cite{huisken} who showed
that the Hawking mass remained monotonic under 
a discontinuous version of the IMC flow.

The motivation behind the present work is to 
explore whether the Ricci flow could lead to a new proof of the
time symmetric Penrose inequality. 
The general Penrose inequality is still an open conjecture and new lines of attack (even in the 
time symmetric case) are certainly of interest. 
Is it possible that a smooth combination of the Ricci flow plus diffeomorphisms
exists so that the evolution
of Hawking mass is monotonic? 
The hope underlying the use of the RF is 
that the discontinuities (like those of the IMC flow) will be smoothed out 
by the RF term. 
There are known examples in fluid mechanics where discontinuities of inviscid flow are smoothed out
by viscous effects. 
With this motivation of developing a
Ricci flow approach to the Penrose inequality, we have made a beginning in 
this paper by studying the evolution of area and Hawking mass under the 
Ricci flow.


In section II we give a brief review of the Ricci flow. 
For a more complete and rigorous account the reader is referred to the 
mathematical literature\cite{topping,rfbook}.
Section III lists the geometrical quantities of physical interest
in this paper. Section IV
paves the way for the general treament by treating the special 
case of spherical symmetry. 
This special case is useful since one can
explicitly work out geometrical quantities of interest and 
develop a physical feel for them. Spherical symmetry is
a good source of physical examples and counterexamples which
guide the general study.
In Section V we give up spherical symmetry and treat the 
evolution of area of a closed surface under the Ricci flow
and derive an inequality relating the rate of change of area and the
Hawking mass. 
Section VI treats the evolution of Hawking mass
under Ricci flow and puts forward a conjectured maximum principle
which governs the long time behaviour of the flow.
Section VII is a concluding discussion. Our 
metric conventions are from Ref.\cite{poisson}.

\section{The Ricci Flow}

Let $(\Sigma,h_{ab})$ be an asymptotically flat, three dimensional Riemannian manifold. 
($a,b$ run over 1,2,3. We restrict our discussion to three dimensional 
manifolds.) Our interest is in asymptotically flat spaces since we are
interested in the energy and entropy of black holes. 
The definitions
of energy, entropy and black holes all need an asymptotic region.
The total energy or ADM mass of an initial data set is only well 
defined if an asymptotic structure (either flat or AdS) is fixed. Black holes 
are defined as regions
of spacetime from which escape to infinity is impossible and thus 
refer to an asymptotic structure.
We require that the metric tend
to a {\it fixed} flat metric $\delta_{ab}$ at infinity 
$h_{ab}\rightarrow\delta_{ab}+O(1/r)$. Given an initial 
metric $h_{ab}$, the Ricci flow evolves the metric according to its Ricci 
tensor. The evolution
parameter is $\tau$ and the family of metrics on $(\Sigma$, 
$h_{ab}(\tau))$
satisfies the Ricci flow equation
\begin{equation}
\frac{\partial{h_{ab}}}{\partial \tau}=-2 R_{ab} .
\label{ricciflow}
\end{equation}
In the neighborhood of a point $p\in \Sigma$, we can introduce 
a Riemann normal co-ordinate system and then the form of 
(\ref{ricciflow}) becomes parabolic ($\nabla^2$ is the Laplacian in
local co-ordinates)
\begin{equation}
\frac{\partial{h_{ab}}}{\partial \tau}= \nabla^2 h_{ab}
\label{heat}
\end{equation} 
and looks like a heat equation for the metric coefficients. However,
in a general co-ordinate system, the PDE (\ref{ricciflow}) is a {\it 
degenerate}
parabolic equation, because of its diffeomorphism invariance.

More generally, we will be interested in the Ricci flow modified
by a diffeomorphism
\begin{equation}
\frac{\partial{h_{ab}}}{\partial \tau}= -2R_{ab}+D_a \xi_b+D_b \xi_a,
\label{ricciplusdiffeo}
\end{equation} 
where $\xi^a$ is any vector field on $\Sigma$ which vanishes at infinity. 
Calculationally, it
is convenient to consider the two terms separately, defining a {\it pure}
Ricci flow (\ref{ricciflow}) and a {\it pure} diffeo
\begin{equation}
\frac{\partial{h_{ab}}}{\partial \tau}= D_a \xi_b+D_b \xi_a.
\label{diffeo}
\end{equation} 
 In the mathematical literature on Ricci flows (which
deals with compact spaces) the term $\lambda h_{ab}$ is sometimes added 
to the RHS of (\ref{ricciflow}) to define a ``normalised Ricci flow''. 
Such a term is inadmissible in the present physical context as it would
rescale the metric at infinity and violate our asymptotic requirement
that the metric tends to a {\it fixed} flat metric at infinity.

In the standard initial value formulation \cite{poisson} of 
general relativity, the basic variables
are the induced metric $h_{ab}$ on a spatial slice $\Sigma$ in the 
space-time manifold $M$,
$\Sigma\in M $ and the extrinsic curvature $k_{ab}$ of $\Sigma$.
We also use the notation $k=h^{ab}k_{ab}$.
These variables are subject to constraints:
\begin{equation}
D_b(k^{ab}-h^{ab}k)=8\pi j^b
\label{constraint1}
\end{equation}
and
\begin{equation}
R+k^2-k_{ab}k^{ab}=16\pi\rho,
\label{constraint2}
\end{equation}
where $j^b$ is the matter current and $\rho$ is the matter density. 
The matter is required to satisfy ``energy conditions'',
the dominant, weak or strong energy condition.

A three dimensional manifold $(\Sigma,h_{ab})$ can be viewed
as a time symmetric intitial data set for Einstein's equations. By ``time
symmetric'' we mean that the extrinsic curvature $k_{ab}$ of 
$\Sigma$ has been set to zero, so that $(\Sigma,h_{ab})$ is totally
geodesic. (This 
is similar to choosing initial data in
classical mechanics so that all the momenta vanish. Dropping
N particles from rest is an example.) 
With $k_{ab}$ (and the matter current $j^a$) set to zero, the 
diffeomorphism constraint (\ref{constraint1}) is automatically satisfied
and the Hamiltonian constraint reduces to $R=16\pi\rho$.
A physically 
important constraint on initial data for general relativity is that 
the data are subject to an energy condition. The dominant,
weak and strong energy conditions all imply the local energy
condition, which states that the local energy density is 
non-negative. 
This
translates into the geometrical statement that the scalar curvature
$R$ of $h_{ab}$ be non-negative. The Ricci flow has the appealing
property that it preserves the non-negativity of scalar 
curvature\cite{rfbook}. 
This follows from the ``maximum principle'' for the scalar curvature.
Under Ricci flow(\ref{ricciplusdiffeo}), the scalar curvature satisfies
the non linear heat type equation
\begin{equation}
\frac{\partial R}{\partial \tau}=\nabla^2 R+2R^{ab}R_{ab}+{\cal L}_\xi R,
\label{scalarequation}
\end{equation}
which implies that $\frac{\partial R}{\partial\tau}$ is positive
at a minimum of $R$.
Eq.(\ref{ricciplusdiffeo}) provides us with a flow on the space
of initial data to Einstein's equations which remains within physically
allowed (non-negative scalar curvature) data.

\section{Geometric Quantities of Interest}
Let $S$ be a closed surface in $\Sigma$, $\gamma_{ij}, (i,j=1,2)$ the pull 
back
or induced metric on $S$, ${\cal R}$ the scalar curvature 
of $(S,\gamma_{ij})$ and
$K$ the trace of its extrinsic curvature. 
We will be interested in the evolution of some geometric 
properties of $S$ under the Ricci flow. These are 
the area of $S$,
\begin{equation}
{\cal A}(S)=\int_S dA =\int_S d^2x \sqrt{\gamma} 
\label{area} 
\end{equation}
and the Hawking mass of $S$
\begin{equation}
{\cal M}_H(S)= 
\frac{\sqrt{{\cal A}(S)}}{64\pi^{3/2}} \int_S dA (2{\cal R} -K^2).
\label{mass}
\end{equation}
Our interest in these quantities 
stems from their physical significance. 
The area of apparent horizons is
related to the entropy of Black Holes and the Hawking Mass is
related to the Energy. 
The Hawking Mass of a surface $S$ can be physically interpreted as 
the mass contained within the surface $S$. While there are some
problems with this interpretation (positivity is not always assured),
the Hawking mass is an useful notion\cite{jang,geroch} of quasilocal mass.
It vanishes in the limit that $S$ shrinks to a point and 
becomes the ADM energy for a round sphere at infinity. In fact,
the supremum of ${\cal M}(S)$ over $S$ is the ADM mass as one can
see from \cite{jang,geroch}.  
Unlike the ADM energy, which is only well-defined for asymptotic
spheres, the Hawking mass is defined for any closed surface $S$.
Under the RF, a surface $S$, which is initially asymptotic
may shrink into the interior of $\Sigma$. For this reason Hawking
mass is a more convenient object to study than the ADM mass.

Rather than the Hawking mass, it is more convenient to deal with 
the related dimensionless quantity
the ``compactness'' of $S$ 
\begin{equation}
{\cal C}(S)=\int_S dA (2{\cal R} -K^2),
\label{compactness}
\end{equation}
which is a combination of the Hawking mass and the area\cite{compact}.
The quantity ${\cal C}(S)$ has been used to 
good effect by Geroch, Jang and Wald \cite{jang,geroch} in their approach 
to 
positive mass theorem and the Penrose inequality. In fact their
work forms the base for recent progress \cite{huisken}
on the Riemannian Penrose inequality. We will see that 
${\cal C}(S)$ tends to zero as $S$ 
tends to a  round sphere of infinitesimal radius and also as $S$ 
tends to an asymptotic round sphere.

\section{Spherical Symmetry}

The Ricci flow is a tensor evolution equation and therefore commutes with diffeomorphisms. 
It follows that isometry groups are preserved under the Ricci flow. One way to 
approach the Ricci flow is to start with symmetric situations so that the complexity of the flow
is reduced.
If we impose so much symmetry that the spaces of interest are homogeneous, the RF becomes 
an ODE rather than a PDE. Such situations have been studied\cite{rfbook}. However, this assumption 
is too restrictive from our present physical motivation. We would like to deal with asymptotically flat
(or AdS ) spaces since energy is defined with respect to an asymptotic structure. Homogeneous 
spaces which are asymptotically flat would be everywhere flat and not very interesting.
Spherical symmetry has the advantage that there {\it are} non trivial 
asymptotically flat spaces.
The Ricci flow reduces to a PDE with just two independent variables, the 
$\tau$ and $r$ co-ordinates.
As we will see, spherically symmetric spaces 
will provide us with an analytical as well as numerical testing ground and pave the way for the 
general treatment. 
This special case is useful since one can develop a physical feel for the geometrical quantities of interest and
easily produce physical examples and counterexamples as a guide to intuition.

While setting up the spherically symmetric initial data set, 
we will work with two forms of the metric. 
Each of these has its use and its limitation. We will refer to them 
as ``a-form'' and ``b-form''. They correspond to different choices of 
co-ordinate gauge.
\begin{itemize}
 \item \textbf{The a-form:} In this form of the initial data set, 
the 3-metric is taken as
   \begin{equation}
    ds^2=a(r)dr^2+r^2(d\theta^2+sin^2\theta d\phi^2).
\label{aform}
   \end{equation}
With this form of the metric we calculate the Ricci tensor and the 
scalar curvature:
The nonzero components of the Ricci tensors are 
(a prime means 
differentiation with respect to $r$)
\begin{equation}
 R_{rr}=\frac{a'}{ra}\nonumber,
\end{equation}
\begin{equation}
 R_{\theta \theta}=\frac{a'r}{2a^2}+1-\frac{1}{a}\nonumber,
\end{equation}
\begin{equation}
 R_{\phi\phi}=sin^2\theta(R_{\theta \theta})\nonumber.
\end{equation}
The scalar curvature is 
\begin{equation}
 R=\frac{2}{r^2}+\frac{2a'}{ra^2}-\frac{2}{ar^2}\nonumber.
\end{equation}

  The a-form is useful for the study of the evolution of Hawking mass ${\cal M}_H$. 
However, this form of the metric is not useful if there is an apparent 
horizon because in 
that case $a(r)$ blows up at the apparent horizon.
To study the evolution of the area of the apparent horizon 
under RF we use the ``b-form'' of the metric discussed next.
\item \textbf{The b-form:} In this form of the initial 
data set, the 3-metric is taken as
   \begin{equation}
    ds^2=dr^2+b(r)(d\theta^2+sin^2\theta d\phi^2).
\label{bform}
   \end{equation}
With this metric we again calculate the Ricci tensor and the 
scalar curvature:
The nonzero components of the Ricci tensor are
\begin{equation}
 R_{rr}=\frac{b'^2-2bb''}{2b^2},\nonumber
\end{equation}
\begin{equation}
 R_{\theta \theta}=1-\frac{b''}{2},\nonumber
\end{equation}
\begin{equation}
 R_{\phi\phi}=sin^2\theta(R_{\theta \theta}).\nonumber
\end{equation}
The scalar curvature is 
\begin{equation}
 R=\frac{b'^2-4b(b''-1)}{2b^2}.
\label{scalarss}
\end{equation}
\end{itemize}
{\it a-form and Hawking Mass:}
Let us start with the a-form (\ref{aform}) of the metric and 
evaluate the Hawking mass functional (\ref{mass})
for $S$ chosen to be a sphere $r=constant$. For this spherical topology,
the first term in (\ref{compactness}) gives $16\pi$ by the Gauss-Bonnet
theorem and  
\begin{equation}
{\cal M}_H(S):=\frac{\sqrt{{\cal 
A}}}{64\pi^{\frac{3}{2}}}\bigg(16\pi-\int_S K^2 
dA\bigg).
\label{hmf}
\end{equation}
Let ${\hat n}_a$ be a unit normal to the surface $S$. The normalization 
\begin{equation}
h^{ab}{\hat n}_a{\hat n}_b=a^{-1}(r){\hat n}_r{\hat n}_r=1 
\end{equation}
fixes 
\begin{equation}
 {\hat n}_r=\sqrt{a}
\end{equation}
and so 
\begin{equation}
 {\hat n}^a=(\frac{1}{\sqrt{a}},0,0).
\end{equation}
The trace of the extrinsic curvature is 
\begin{equation}
K=D_a{\hat n}^a=\frac{2}{r\sqrt{a}},
\end{equation}
so we have
\begin{equation}
 \int K^2 dA=\frac{16\pi}{a(r)},
\label{extrinsicss}
\end{equation}
and 
\begin{equation}
 {\cal M}_H(S)=\frac{\sqrt{\cal 
A}}{4\pi^{1/2}}\bigg[1-\frac{1}{a(r)}\bigg].
\label{hawkingss}
\end{equation}
For flat space $a(r)=1$ and we have $M_H(S)=0$ for round spheres in 
flat space as expected.
If we take the example of the exterior Schwarzschild space
\begin{equation}
 a(r)=\bigg(1-\frac{2M}{r}\bigg)^{-1}
\end{equation}
and then for any $r>2M$,
\begin{equation}
 M_H(S)=\frac{\sqrt{4\pi 
r^2}}{4\sqrt{\pi}}\bigg[1-\bigg(1-\frac{2M}{r}\bigg)\bigg]=M
\end{equation}
For a general a-form metric we can 
write $a(r)$ 
as $\bigg(1-\frac{2M(r)}{r}\bigg)^{-1}$. 
We find that\cite{poisson}
\begin{equation}
 {\cal M}_H(S)=M(r)
\end{equation}
and the ``compactness'' works out to 
\begin{equation}
 C(r)=32\pi \frac{M(r)}{r}.
\label{sscofr}
\end{equation}

\begin{figure} 
\centering \includegraphics[width=100mm]{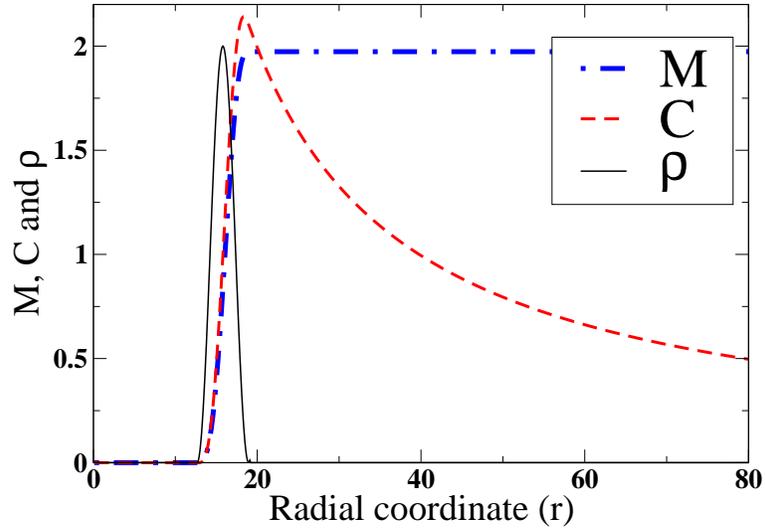} 
\caption{\label{mrhoc4} Note that 
${\cal M}_H(r)$ increases with $r$ to attain its 
asymptotic ADM value. 
But $C(r)$ increases to a maximum value and then
decreases to $0$ at infinity.} 
\end{figure}

\begin{figure}
\centering
\includegraphics[width=100mm]{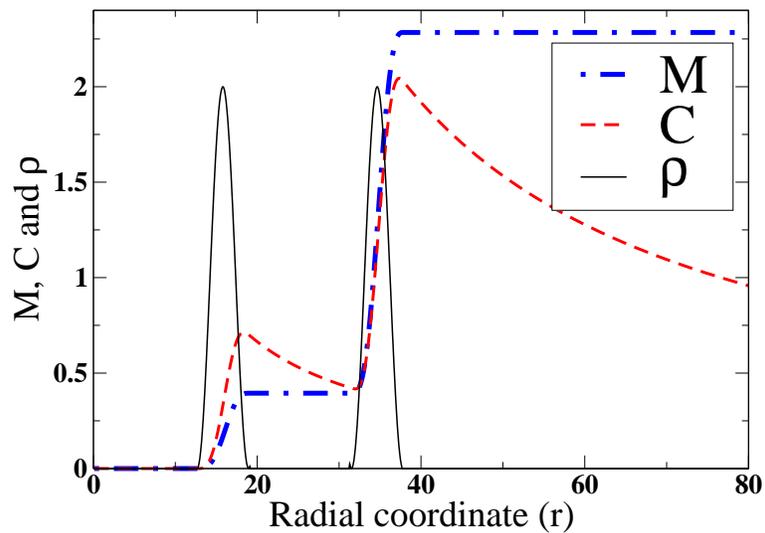}
\caption{\label{plot4} Two shells of matter.}
\end{figure}

In order to get a better physical feel for what these geometrical 
quantities mean, let us consider some typical distributions of matter. Let 
us choose the matter density  $\rho(r)=(1/16\pi)R(r)$ positive
and plot the functions $\rho(r)$, $C(r)$ and 
$M_H(r)$.  Figure(\ref{mrhoc4}) displays the forms of these functions
(in arbitrary units) for a spherical shell of matter. 
$C(r)$ increases to a maximum value
and then decreases to $0$ at infinity.  For a matter 
distribution of two momentarily static shells of matter, 
the slightly more complex behaviour of $C(r)$ is shown in  
Fig. (\ref{plot4}). For a star $C(r)$ 
attains its maximum near the surface of the star. 
From (\ref{sscofr}), in the Newtonian limit $C(r)$ is a 
constant times the dimensionless Newtonian potential, or the mass to 
radius ratio. Hence the name ``compactness'' is justified.

Note that $M_H(r)$ monotonically increases with $r$ to attain
its asymptotic value. 
This is due to the local energy condition, which implies 
positive scalar curvature $R\ge0$. This condition is conveniently 
stated in terms of the function $M(r)$. 
The scalar curvature of (\ref{aform}) is given by
$R=4M'(r)/r^2$ and so the constraint of positivity
of scalar curvature simply states that $M(r)$ is a 
non-decreasing function of $r$. 
Assuming that the form (\ref{aform})
holds all the way to the origin and that the scalar 
curvature $R$ is finite, we have $M(0)=0,M'(0)=0, M''=0, M'''(0)\ge0$. 
$M(r)$ increases from zero and tends to an asymptotic value 
$M_{ADM}=\lim_{r->\infty} M(r)$ which is the ADM mass of the metric. 
In a sense\cite{poisson}, 
${\cal M}_H(r)$ measures the total mass 
contained within a sphere of areal radius $r$. ${\cal M}_H(r)$ is 
non-negative
for all $r$ and 
and from (\ref{hawkingss}), we conclude that $a(r)\ge1$ for all $r$.

Note that from (\ref{extrinsicss}) it follows that if the space
contains an apparent horizon ($K=0$), $a(r)$ must diverge. This
is why the a-form is unsuitable for treating apparent horizons. 
The b-form does not suffer from this problem.

{\it Geometric quantities in the b-form:}
We now consider a round sphere $S\subset \Sigma$ given by 
$r=constant$ in the spherically symmetric ``b-form'' (\ref{bform}) of the 
metric.
The unit normal satisfies
\begin{equation}
h^{ab}{\hat n}_a{\hat n}_b={\hat n}_r{\hat n}_r=1 
\end{equation}
and $\hat{n}^r=(1,0,0)$. The area ${\cal A}$ of $S$ is given as
\begin{equation}
 {\cal A}(r)=\int_S \sqrt{\gamma} d\theta d\phi=4\pi b(r),
\label{areaah}
\end{equation}
where $\gamma=b^2 sin^2\theta$ is the determinant of the induced metric $\gamma_{ij}$ on $S$.
The trace of the extrinsic curvature is given
by 
\begin{equation}
 K:=D_a{\hat n}^a=\frac{b'}{b},
\label{ex}
\end{equation}
(where a prime indicates differentiation with respect to $r$).
The general formula for the compactness reduces in spherical symmetry to,
\begin{equation}
 C(r)=16\pi-\int_S \sqrt{\gamma} d\theta d\phi K^2=16\pi-\frac{4\pi 
b'^2}{b}
\label{compactss}
\end{equation}
and the Hawking mass is 
\begin{equation}
 {\cal 
M}_H(r)=\frac{\sqrt{{\cal 
A}(r)}}{64\pi^{3/2}}C(r)=\frac{\sqrt{b}}{2}[1-\frac{b'^2}{4b}].
\label{hawkingss2}
\end{equation}
Note that ${\cal M}_H$ depends on $b(r)$ and its derivative $b'(r)$,
in constrast to the simple algebraic relation (\ref{hawkingss}) we had
in the a-form of the metric.

{\it Area under Ricci flow:} 
We consider a metric initially in the b-form, evolving under 
a pure Ricci flow (without a diffeomorphism term). Under this evolution, 
the b-form may not be preserved. We view 
$S$ as a fixed surface of $\Sigma$ and so the co-ordinate location 
of the 
surface $S$ does not change and hence $\frac{dr}{d\tau}=0$.
From the Ricci flow we have
\begin{equation}
 \frac{\partial h_{\theta \theta}}{\partial \tau}=-2R_{\theta \theta} \nonumber
\end{equation} 
from which follows the formula
\begin{equation}
 \frac{d{\cal A}}{d\tau}=4\pi \frac{\partial b}{\partial \tau}=-4\pi 
(2-b'')\nonumber
\end{equation}
for the {\it instantaneous} rate of change of area of $S$.

Using the  scalar curvature $R$ for the ``b-form'' of the metric 
(\ref{scalarss})
we see that, in spherical symmetry,
\begin{equation}
 \frac{d{\cal A}}{d\tau}=-\frac{1}{2}\int_S \sqrt{\gamma} d\theta d\phi R 
- \frac{1}{4} C
\label{dadtss}
\end{equation}
and so we arrive at the inequality (since $R\ge0$)

\begin{equation}
 \frac{d{\cal A}}{d\tau}\le - \frac{1}{4} C.
\label{dadtineq}
\end{equation}
In the case of the Schwarzschild space, $R=0$ and so the first integral
in (\ref{dadtss}) vanishes and we have
\begin{equation}
\frac{d{\cal A}}{d\tau}= -\frac{1}{4}C(S).
\label{dadtsch} 
\end{equation} 
Thus Schwarzschild space saturates our inequality (\ref{dadtineq}),
just as it saturates the Penrose inequality. 

{\it Area of apparent horizons under Ricci Flow:}
Let $S$ be a minimal surface (or apparent horizon, they coincide 
in the case of time symmetric data) in 
$\Sigma$ {\it i.e}  $S$ is a closed two 
manifold embedded in $\Sigma$ with the property that the 
trace of the extrinsic curvature vanishes. 
We want to see how the area of $S$ varies under the RF.
We start with the spherically symmetric ``b-form'' of the 
metric (\ref{bform}).
Let the location of the apparent horizon be at $r=r_0$. From (\ref{ex}),
we have that $K=b'/b|_{r_0}=0$
The condition that the surface $r=r_0$ be an apparent 
horizon is
\begin{equation}
 b'|_{r=r_0}=0.
\label{ahc}
\end{equation}

Can a minimal surface spontaneously appear if none was present
initially? The answer is no, as the following argument shows by 
contradiction. Let us 
consider the $b$ form of the metric and evolve it by the Ricci flow 
supplemented by a suitable radial diffeomorphism $\xi^a$  
chosen to maintain the $b$-form
\begin{equation}
\frac{\partial h_{rr}}{\partial \tau} = -2R_{rr} + 2(D\xi)_{rr} = 0, 
\label{rr eqn}
\end{equation}

\begin{equation}
\frac{\partial h_{\theta\theta}}{\partial \tau} = \frac{\partial 
b}{\partial \tau} =
-2R_{\theta\theta} + 2(D\xi)_{\theta\theta}. \label{tt eqn}
\end{equation}

We may write $\xi_{r} = \partial_{r}f$ and using (\ref{rr eqn}) and choose 
$f$ so that
\begin{eqnarray*}
f^{\prime\prime} = \frac{b^{\prime 2} - 2bb^{\prime\prime}}{2b^{2}}.\nonumber
\end{eqnarray*}
(This leaves some freedom in $f$, but this does not affect the following.)
(\ref{tt eqn}) then gives us the evolution equation for $b(r,\tau)$ in the 
$b$ form of the metric
\begin{eqnarray*}
\frac{\partial b}{\partial \tau} = b^{\prime\prime} - 2 + b^{\prime} 
f^{\prime} \nonumber.
\end{eqnarray*}
Differentiating this equation we arrive at an evolution equation 
for $b'$
\begin{equation}
\frac{\partial b^{\prime}}{\partial \tau} = b^{\prime\prime\prime} + 
b^{\prime\prime} f^{\prime} + b^{\prime} f^{\prime\prime},
\label{bprevolve} 
\end{equation}
which gives us a maximum principle for $b'$.
Suppose that $b' > 0$ for all $\tau < \tau_{0}$ and that 
for the first time $\tau = \tau_{0}$, a minimal surface 
appears $(b'(r_{0}) = 0)$ at $r = r_{0}$.
We have $\frac{\partial b'}{\partial \tau} (r_{0}) < 0$, 
since $b'$
decreased to zero. On the other hand, since $r_{0}$ is a minimum of 
$b'$, we have $b'' (r_{0}) = 0$ and 
$b''' (r_{0}) \geq 0$. A glance at (\ref{bprevolve}) 
shows
that $\frac{\partial b'}{\partial \tau} \geq 0$, which is
a contradiction. Thus a minimal surface cannot spontaneously appear under
RF if it was not initially present.

Regions where $b' < 0$ are called trapped regions. Since
the RF is continuous, such regions evolve continuously. Trapped regions are
bounded by minimal surfaces $(b' = 0)$. From the last paragraph 
it is clear that trapped regions cannot spontaneously appear under RF 
since they are accompanied by minimal surfaces. However, trapped 
regions can 
continuously shrink to zero and disappear. When this happens, the two
minimal surfaces that form the boundary of the trapped region merge and
disappear. More descriptively, a minimal surface 
$(b' = 0,\;\;b'' > 0 )$ merges with a maximal surface 
$(b' = 0,\;\;b''< 0 )$ and disappears.

Setting aside such mergers, let us study how the area of minimal surfaces 
evolves under the pure RF (1).
During the RF the metric changes and the location of the horizon may
change and 
so $r_0=r_0(\tau)$ where $\tau$ is the parameter of the RF. 
Also the geometry of $S$ will change. In principle both these effects 
could lead to change of area.
The area is given by (\ref{areaah}) ${\cal A}(r)=4\pi b(r)$.
The total rate of change of area, therefore, is 
\begin{equation}
 4\pi \frac{db}{d\tau}=4\pi\bigg[\frac{\partial b}{\partial 
r}\bigg|_{r=r_0} \frac{dr_0}{d\tau}+\frac{\partial b}{\partial \tau}\bigg|_{r=r_0}\bigg].
\label{adot4}
\end{equation}
The first term vanishes because of the apparent horizon 
condition (\ref{ahc}). The second term is evaluated by 
specialising
eq.(\ref{dadtineq}) to an apparent horizon.
So the area ${\cal A}=4\pi b$ satisfies
\begin{equation}
 \frac{\partial {\cal A}}{\partial \tau}\le-4\pi\nonumber. 
\end{equation}
This implies that the area of the horizon is decreasing at least 
linearly with $\tau$. 
Since the area was finite to begin with, we find that (if the horizon 
persists) $b$ 
evaluated at the horizon goes to zero in a finite $\tau$.

Next we see that as $b\to 0$ we approach a singularity. To show this we will suppose that $R$ is finite and will arrive at a contradiction.
As $b\to 0$  the scalar curvature
\begin{equation}
 R=\frac{2(1-b'')}{b}\nonumber
\end{equation}
is finite only if $b''=1$. We Taylor expand in powers of $r-r_0$ about $r_0$, the location of the apparent horizon
\begin{equation}
 b(r)=b(r_0)+b'(r_0)(r-r_0)+\frac{1}{2}b''(r_0)(r-r_0)^2+...\nonumber.
\end{equation}
As $b(r_0)\to 0$, $b''(r_0)\to1$ and $b'(r_0)=0$ due to the apparent 
horizon 
condition (\ref{ahc}), we have
\begin{equation}
 b(r)=\frac{1}{2}(r-r_0)^2\nonumber,
\end{equation}
so the metric is
\begin{equation}
    ds^2=dr^2+\frac{1}{2}(r-r_0)^2(d\theta^2+sin^2\theta 
d\phi^2)\nonumber.
\end{equation}
Shifting the $r$ co-ordinate $r\to(r-r_0)$ gives the form
\begin{equation}
    ds^2=dr^2+\frac{1}{2}r^2(d\theta^2+sin^2\theta d\phi^2).\nonumber
\end{equation}
So the volume of a ball of radius $r$ is
\begin{equation}
\int d\theta d\phi dr \frac{\sqrt{r^4 
sin^2\theta}}{4}=\frac{1}{4}\bigg(\frac{4\pi r^3}{3}\bigg). \nonumber
\end{equation}
We then find from the expression for the volume of a ball of radius $r$ 
(as $r \to 0$) centered at point $p$ that
\begin{equation}
 volume ~~\mathcal{B}(p,r)=\bigg(\frac{4\pi}{3}\bigg)\bigg(r^3-\frac{1}{30}R(p)r^5 \bigg)\nonumber
\end{equation}
and so the scalar curvature blows up, $R(p)\sim r^{-2}$,  which is a 
contradiction to the assumption of finite $R$ that we started with.
So an apparent horizon which persists under RF results in a singularity in 
a finite amount of $\tau$ parameter $\tau\le \tau_0$, 
where $\tau_0={\cal A}(0)/(4\pi)$.

{\it Compactness under RF:}
We have previously derived the relation (\ref{sscofr}) for the compactness
of a sphere $r=const$ from which follows
\begin{equation}
C(r)=16\pi[1-1/a].
\label{carelation}
\end{equation}
A calculation reveals that under RF
\begin{equation}
\frac{dC}{d\tau}=-\frac{2M'(r)}{r^2}+\frac{2M''}{r}.
\label{compactdotss}
\end{equation}
While the first term here is of definite sign, the second term is not.
As  a result, the rate of change of $C$ with $\tau$ is
not monotonic. Note however that for Schwarzschild $M(r)$ is a constant
$M$ and so $dC/d\tau=0$.

The Hawking mass is given by ${\cal M}_H(r)=rC(r)/(32\pi)$ and to work
out its rate of change $\frac{d{\cal M}_H}{d\tau}$ under the RF, we need 
to find $dr/d\tau$. This is
easily read off from (\ref{dadtss}) and we find
\begin{equation}
\frac{dr}{d\tau}=\frac{-M'}{r}-\frac{C}{32\pi r}.
\label{drdt}
\end{equation}
Putting these equations together we can work out $\frac{d{\cal 
M}_H}{d\tau}$ and it turns out to be a linear combination of $M'$ and $M''$.
As a result, the Hawking mass is {\it not monotonic} along the pure Ricci 
flow in
the a-form of the metric.
It may be that adding a suitable diffeomorphism will result in monotonic 
behaviour for ${\cal M}_H$.  

{\it Maximum principle for compactness:}
However, we can make a diffeo invariant statement about the behaviour
of compactness. We have already seen that if a minimal surface is
initially present this may result in a finite $\tau$ singularity. Let us
suppose that no minimal surface is initially present. None develops under
RF, as we saw earlier. We can therefore use the a-form of the metric. We 
have seen that
$C(r)$ starts from $0$ at $r=0$, reaches a maximum (perhaps several local
maxima) and then decays to zero as $r\to\infty$. Let us focus on 
$C_{max}$ the value of $C$ at its absolute maximum. Just as for apparent
horizons, the value of $C_{max}$ is not affected by a diffeomorphism:
moving the surface does not affect the value of $C_{max}$ since we are at 
a maximum. We choose a diffeomorphism to preserve the a-form of the metric
and then the evolution equation for $a(r)$ reads

\begin{equation}
\frac{\partial a(r)}{\partial \tau}=\frac{a''(r)}{a(r)}-
\frac{3(a'(r))^2}{2a(r)^2}-\frac{2(a(r)-1)+ra'(r)(1-a(r))/a(r)}{r^2},
\label{coeffdot4} 
\end{equation}
where a prime denotes differentiation with respect to $r$. We have studied 
the evolution of (\ref{coeffdot4}) using numerical techniques. Some of the
results presented here were initially suggested by the numerical evidence.

Let us focus on the maximum value of $a(r)$. Recall that $a(r)\ge1$.
At the maximum value of $a(r)$, 
we have $a'(r)|_{max}=0$ and $a''(r)|_{max}\le0$. 
So from equation (\ref{coeffdot4}) we see that the maximum value of 
$a(r)$ is monotone non-increasing as the flow parameter $\tau$ increases 
i.e.,
\begin{equation}
\frac{\partial a(r)_{max}}{\partial \tau}= 
\frac{a''(r)}{a(r)}-\frac{2(a(r)_{max}-1)}{r^2}\le0.
\label{adotineq4} 
\end{equation}
A maximum of $C$ corresponds via (\ref{carelation}) to a maximum of
$a(r)$.  Clearly, a maximum principle for $a(r)$ 
implies a maximum principle for compactness $C(r)$
and we have the inequality
\begin{equation}
 \frac{dC(r)_{max}}{d\tau}\le 0.
\label{cmaxdot4}
\end{equation}
We can use this maximum principle to comment on the long time existence
of the spherically symmetric asymptotically flat  Ricci flow. 
Let us suppose that no minimal surface is initially present. From 
(\ref{adotineq4}), we see that 
for the LHS of (\ref{adotineq4}) to vanish we must have $a(r)_{max}=1$, 
which implies $a(r)=1$ identically. This describes flat space. If the 
initial metric is not flat space, its $a(r)_{max}$ decreases with the
flow and finally attains the flat space fixed point $a(r)=1$.
Our argument shows clearly that in spherical symmetry, the only asymptotically flat
fixed point of the flow is flat space. This recovers results 
obtained by other methods \cite{eric3,ivey}.

To summarise, the maximum principle for compactness leads to a 
criterion for the existence of the Ricci flow in the asymptotically
flat case. If there are no apparent horizons, the flow exists for all 
$\tau$ and converges to flat space. If there are apparent horizons, the
Ricci flow either terminates in finite time singularity or removes the
horizons by mergers.

\section{Area Under Ricci Flow}
We now drop the assumption of spherical symmetry and deal with 
a general asymptotically flat manifold $\Sigma$, with one end at infinity
and a fixed closed orientable surface $S$ of arbitrary topology
embedded in $\Sigma$. The induced metric of $S$ is written 
$\gamma_{ij}$ and extrinsic
curvature tensor of $S$ is written $K_{ij}$, where $i,j$ are two 
dimensional indices in the space tangential to $S$. We will sometimes
use projected $a,b$ indices for these. 
The trace 
of the extrinsic curvature is written $K=\gamma^{ij}K_{ij}$.
As the metric
$h_{ab}$ evolves according to the Ricci flow (\ref{ricciflow}), how 
does 
the
area change with $\tau$? 
Since the metric of $\Sigma$ is changing, we have
to remember that the unit normal to $S$ is also changing with the metric.

Let us define the surface $S$ as the level set of  a function $\eta$ on 
$\Sigma$ which is strictly increasing outward from $S$. 
Quite independent of any metric, the normal $\eta_a=D_a \eta$ is 
a well defined co-vector. $\eta_a$ is non-zero (since we assumed $\eta$
is not locally constant). (Choosing a different function $\eta$ will result in 
multiplication of $\eta_a$ by a positive function on $S$.) The unit normal
\begin{equation}
\hat{n}_a=\frac{\eta_a}{(\eta.\eta)^{1/2}}
\label{unit}
\end{equation}
depends on the metric. As the metric changes the unit normal can only 
change by a multiple of itself $d{\hat n}_a/dt=\alpha {\hat n}_a$.
By differentiating ${\hat n}.{\hat n}=1$, we arrive at 
\begin{equation}
\alpha=\frac{1}{2}\frac{dh^{ab}}{d\tau} \hat{n}_a \hat{n}_b,
\label{alpha}
\end{equation} 
(where in our notation, $\frac{dh^{ab}}{d\tau}$ is {\it defined} as 
$\frac{dh_{ab}}{d\tau}$ with its indices raised using the metric 
tensor $h^{ab}$).

Starting from the definition (\ref{area}) we compute 
$\frac{d{\cal A}}{d\tau}$
\begin{equation}
\frac{d{\cal A}}{d\tau}=\int_S d^2 x \frac{d\sqrt{\gamma}}{d\tau}.
\label{dadt1}
\end{equation} 
We easily see that 
\begin{equation}
\frac{d\sqrt{\gamma}}{d\tau}=
\frac{1}{2}\sqrt{\gamma}\gamma^{ab}\frac{dh_{ab}}{d\tau}
\label{rootgammadot}
\end{equation}
and we arrive at 
\begin{equation}
\frac{d{\cal A}}{d\tau}=\frac{1}{2}\int_S \sqrt{\gamma}d^2 x 
(h^{ab}-\hat{n}^a\hat{n}^b)\frac{dh_{ab}}{d\tau}.
\label{dadt2}
\end{equation} 

{\it Area under ricci flow:}
Using the form (\ref{ricciflow}) of the Ricci flow, we find
\begin{equation}
\frac{d{\cal A}}{d\tau}=\int_S \sqrt{\gamma} d^2 x 
[\hat{n}^a \hat{n}^b R_{ab}-R].
\label{dadt3}
\end{equation} 
From the Gauss-Codazzi equation\cite{poisson}, we
have  
\begin{equation}
\hat{n}^a \hat{n}^b R_{ab}-R=-1/2[R+{\cal 
R}+(K_{ij}K^{ij}-K^2)],
\label{gausscodazzi}
\end{equation} 
which can be rearranged to give 
\begin{eqnarray} 
\nonumber
\frac{d{\cal A}}{d\tau}&=&
-1/2\int_Sd^2 x \sqrt{\gamma} \bigg[R+(K^{ij}-\frac{1}{2}K 
\gamma^{ij})(K_{ij}-\frac{1}{2}K 
\gamma_{ij})\bigg]\\ 
&-&\frac{1}{4}\int_Sd^2 x\sqrt{\gamma} (2{\cal R} -K^2). 
\label{dadt4p} 
\end{eqnarray} 

The second integral in (\ref{dadt4p}) is 
identified as $-{\cal C}(S)/4$, minus one fourth the compactness 
integral
of $S$ and the first integral, 
which is of definite sign can be dropped to arrive at 
the inequality 
\begin{equation} 
\frac{d{\cal A}}{d\tau}\le -\frac{{\cal C}(S)}{4}.
\label{main} 
\end{equation} 
This inequality is one of the main results of 
this paper. This result can be reexpressed in terms of the Hawking Mass: 
\begin{equation} 
\frac{d{\cal A}^{3/2}}{d\tau}\le -24 \pi^{3/2} {\cal M}_H(S).
\label{mainmassform} 
\end{equation} Thus the rate of decrease of area of a closed 
2-surface under Ricci 
flow is bounded by the Hawking 
mass.
As we mentioned earlier, the inequality (\ref{main}) is saturated
in the case of the spheres of Schwarzschild space (which is given by 
(\ref{aform}) with $a(r)=(1-2M/r)^{-1})$. In this case $R=0$ and the 
spheres are
shear free ($K_{ij}=\frac{1}{2}K\gamma_{ij}$), so the first integral
in (\ref{dadt4p}) vanishes. We conjecture that round spheres in 
Schwarzschild are the only surfaces which saturate this bound.

As a simple application of this inequality, let us consider 
flat space. Since the Ricci tensor vanishes we have that 
$d{\cal A}/d\tau=0$ and so the LHS of (\ref{mainmassform}) vanishes. We 
arrive
at the conclusion that for all surfaces in flat space, the Hawking
mass is non positive! This fact has also been noticed in \cite{hayward},
where a direct proof is given. In fact, the converse of this 
statement is also true: Given positive scalar curvature,
flat space is the only one for which the
Hawking mass is non-positive. To see this, note that the supremum
over $S\in \Sigma$ of the Hawking mass is the ADM mass and if this supremum vanishes,
it follows from the positive mass theorem  that the space must be
flat.

{\it Area under diffeos:}
Under a diffeo, the metric changes as in (\ref{diffeo})
and so we have from (\ref{dadt2})
\begin{equation}
\frac{d{\cal A}}{d\tau}=\frac{1}{2}\int \sqrt{\gamma}d^2x 
(h^{ab}-{\hat n}^a{\hat n}^b)2D_a\xi_b=\int \sqrt{\gamma}d^2x 
[D_a\xi^a-n^an^bD_a\xi_b]
\end{equation}
If we suppose that $\xi^a$ is tangent to $S$, then
with ${\tilde D}_a$ denoting covariant derivative intrinsic 
to $(\gamma_{ab},S)$,
\begin{equation}
\tilde{D}_a\xi_b=\gamma_a^{~a'}\gamma_b^{~b'} D_{a'}\xi_{b'}
\end{equation}
gives us 
\begin{equation}
\tilde{D}_a\xi^a=\gamma^{ab}\tilde{D}_a\xi_b=(h^{ab}-\hat{n}^a\hat{n}^b)D_a\xi_b
 \end{equation}
 So, as one would expect,
 \begin{equation}
  \frac{d{\cal A}}{d\tau}=\int \sqrt{\gamma}d^2x [\tilde{D}_a\xi^a]=0,
 \end{equation}
since this is a divergence over a closed surface $S$.
It is therefore enough to consider the component of $\xi$ normal to $S$. Let 
 \begin{equation}
  \xi^a=u{\hat n}^a.
 \end{equation}
Then
\begin{equation}
  \frac{d{\cal A}}{d\tau}=\int \sqrt{\gamma}d^2x  (h^{ab}-{\hat 
n}^a{\hat n}^b)D_a(u{\hat n}_b) ,
 \end{equation} 
which works out to the standard answer
\begin{equation}
  \frac{d{\cal A}}{d\tau}=\int \sqrt{\gamma}d^2x ~ u\gamma^{ab}D_an_b=\int 
\sqrt{\gamma}d^2x ~ uK .
\label{adotdiffeo6}
\end{equation}

{\it Area of horizons under ricci flow:}
If $S$ is a minimal surface, we find from (\ref{compactness}) that since 
$K=0$, the area changes 
according to 
\begin{equation}
 \frac{d{\cal A}(S)}{d\tau}\le \frac{1}{4}C(S)=-\frac{8\pi\chi(S)}{4},
\end{equation}
where $\chi(S)$ is the Euler characteristic of $S$.
For a minimal surface of spherical topology, we have
\begin{equation}
 \frac{d{\cal A}}{d\tau}\le -4\pi,
\label{sphericalshrink}
\end{equation}
which we had seen earlier for the special 
case of spherical symmetry.
This result is unaffected by adding a diffeo to the RF because 
of (\ref{adotdiffeo6}) and the minimal surface condition $K=0$.

An interesting special case is one for which $S$ is the 
outermost horizon with respect to asymptotic infinity.
S is defined as the boundary of the region having 
trapped surfaces. S is a minimal surface and 
is known to have spherical topology. According to (\ref{sphericalshrink}),
the  area of $S$ will shrink.
Under the Ricci flow, the trapped region cannot disappear 
suddenly, but evolves continuously, because trapped surfaces remain 
trapped 
under small perturbations. 
Near $S$ there will be a new 
minimal surface with the same area. If the initial area is ${\cal A}_0$, 
it dissappears 
within a time ${\cal A}_0/4\pi$, either by merger or by shrinking to zero. 



\section{Hawking Mass under Ricci Flow}
The Hawking mass of a closed surface $S$ given by 
(\ref{mass}) is a combination of the ``compactness'' of $S$ and its 
area. To study the evolution of the Hawking mass, it is enough to
understand the evolution of the compactness, the evolution
of area being already treated in the last section.
From the formula (\ref{compactness}) for compactness, we see that
the first term $\int d^2x\sqrt{\gamma}{\cal R}$ drops out on 
differentiation
since it is a topological invariant by the Gauss-Bonnet theorem. The
second term gives
\begin{equation}
\frac{dC(S)}{d\tau}=
-\frac{d}{d\tau}\int 
K^2\sqrt{\gamma}d^2x=-\int 2K\sqrt{\gamma}\frac{dK}{d\tau}d^2x
-\int K^2\frac{d\sqrt{\gamma}}{d\tau} d^2x.
\label{dcdt1}
\end{equation}
Using (\ref{rootgammadot}) for the second term in (\ref{dcdt1}) and the 
formula 
\begin{equation}
\frac{dK}{d\tau}=\frac{d}{d\tau}(D_a\hat{n}^a)=\frac{d\Gamma^a_{am}}{d\tau}
\hat{n}^m+D_a 
\frac{d\hat{n}^a}{d\tau},
\label{dkdt}
\end{equation}
we find after a straightforward calculation that
\begin{eqnarray}
\frac{dC}{d\tau}&=&
-\int dA K \bigg[ h^{ab}\hat{n}^cD_c 
\frac{dh_{ab}}{d\tau}-2D_a(\frac{dh^{ab}}{d\tau} 
\hat{n}^b)+\hat{n}^aD_a(\frac{dh^{cd}}{d\tau} 
\hat{n}_c\hat{n}_d)\bigg]\\
&-&
\int dA \frac{K^2}{2}\bigg[ \frac{dh^{cd}}{d\tau} 
\hat{n}_c\hat{n}_d+h^{ab}\frac{dh^{ab}}{d\tau} \bigg].
\label{compacgen}
\end{eqnarray}

Equations (\ref{compacgen}) give the general evolution of the compactness under 
for any one parameter family 
$h_{ab}(\tau)$ of metrics. We will specialise these equations to the two cases of interest:
the pure Ricci flow and a pure diffeo.

{\it Compactness under ricci flow:} Substituting 
$\frac{dh_{ab}}{d\tau}=-2R_{ab}$ into (\ref{compacgen}) 
we  find after simplification using the contracted Bianchi identity 
that 
\begin{equation}
\frac{dC}{d\tau}=\int_S 
dA\{K^2(R+\hat{n}^a\hat{n}^bR_{ab})-
2K[2R^{ab}D_a\hat{n}_b-\hat{n}^aD_a(R^{cd}\hat{n}_c\hat{n}_d)]\}.
\label{compacricci}
\end{equation}

{\it Compactness under diffeos:} How does the compactness change under  a 
pure diffeo generated by $\xi^a$? 
As we saw for the area, the tangential component of $\xi^a$ does not cause 
any change in the integral. The
diffeo vector field can be characterised by its normal component
$\xi^a=u \hat{n}^a$. Substituting (\ref{diffeo}) into (\ref{compacgen})
and simplifying and using the Gauss-Codazzi equation \cite{poisson}
\begin{equation}
 -2n^an^bR_{ab}+R=\mathcal{R}+(k^{ij}k_{ij}-k^2)
\end{equation}
gives us the formula \cite{geroch}

\begin{equation}
\frac{dC}{d\tau}=\int_S[2K\tilde{D}^a\tilde{D}_au+uK\sigma^{ij}\sigma_{ij}
+ukR-\frac{1}{2}uK(2\mathcal{R}-K^2)] 
\sqrt{\gamma}d^2x 
\label{cdot3.6}
\end{equation}
where $\tilde{D_a}$ denotes the intrinsic covariant derivative operator within the surface 
and $\sigma^{ij}=K^{ij}-1/2\gamma^{ij}K$.
If there exists a diffeo such that $uK=1$ (this is the inverse mean
curvature flow (IMC)) one can conclude \cite{geroch} that under this 
diffeo
\begin{equation}
\frac{dC}{d\tau}\ge - C/2
\label{geroch}
\end{equation}
This inequality (which too is saturated by the Schwarzschild space) 
was used by Jang and Wald\cite{jang} to prove
the Penrose inequality.
We can get an expression for the evolution of Hawking mass. The Hawking 
mass is
\begin{equation}
 {\cal M}_H(S)= 
\frac{\sqrt{{\cal A}(S)}}{64\pi^{3/2}} C(S).
\label{hm1.6}
\end{equation}
We then have
\begin{equation}
 \frac{d}{d\tau} 
{\cal 
M}_H=\Bigg(\frac{1}{64\pi^{3/2}}\Bigg)\Bigg(\frac{1}{2\sqrt{{\cal 
A}}}\frac{d{\cal A}}{d\tau}C+\sqrt{{\cal A}}\frac{dC}{d\tau}\Bigg) 
\end{equation}
Knowing $\frac{d{\cal A}}{d\tau}$ and $\frac{dC}{d\tau}$ 
for a flow,
can calculate $\frac{d}{d\tau} {\cal M}_H$. (\ref{adotdiffeo6})and 
Geroch's inequality
(\ref{geroch}) then imply that the Hawking mass is monotonic under 
the IMC flow.

{\it A maximum principle for compactness?}
The compactness $C(S)$ of a closed surface $S$ tends to zero for $S$ 
tending to 
a small round sphere, and also tends to zero for 
a round spheres at asymptotic infinity. We would expect based on 
our experience with spherical symmetry, that
somewhere in between, there is a surface $S$ for which
$C(S)$ attains its global maximum. If this maximum is equal to 
$16\pi$, then $S$ is a minimal surface and conversely.
If the maximal compactness is less than $16\pi$, we may
expect from the spherically symmetric case that this value
$C(S)|_{max}$ will monotonically decrease to zero.
\begin{equation}
\frac{dC_{max}}{d\tau}\le0, 
\label{maxpple}
\end{equation}
with the equality holding only for flat space.
A zero value for $C(S)|_{max}$ would also imply a non-positive value for 
${\cal M}_H(S)$. Supremising this over $S$ tells us that the ADM mass of 
the space vanishes, which implies (by the positive energy 
theorem\cite{schoen}) that
space is flat. 
We conjecture that (\ref{maxpple}) is true generally, 
but have not
been able to establish the truth of this conjecture.
If this conjecture is true, it would imply that in the long time,
an initial metric without minimal surfaces would approach flat space.

 \section{Conclusion}

We have described some applications of Ricci flow techniques
to asymptotically flat spaces in general relativity. Our main result is that 
under the Ricci flow, the rate of change of area of a closed 
surface is bounded by its Hawking mass. This inequality 
is saturated by the Schwarzshild space. Since the 
Schwarzschild space saturates 
the Penrose inequality as well as Geroch's (\ref{geroch}), 
our inequality may be related to these.
We have also studied the behaviour of compactness
under the Ricci flow. Our work in spherical symmetry suggests that
there may be a maximum principle for the compactness. The compactness
is a {\it functional} on closed two dimensional surfaces. The conjecture 
is that for positive scalar curvature, the maximum value of this {\it 
functional} decreases under the Ricci flow plus diffeomorphisms.
This proposal is a functional maximum principle unlike the more usual
ones which are which are formulated on functions.
If such a principle
does exist, it would be of great interest as a diagnostic for the long
term existence of the asymptotically flat Ricci flow. 
Most of the 
mathematical work in the area of Ricci flows is concerned with closed
three manifolds. One of our main points here is that there may be 
interesting physical applications of these ideas to asymptotically flat 
spaces. For instance, \cite{eric1} poses the question of stability of 
Euclidean flat space $I\!\!R^3$ under RF. If a maximum principle for 
compactness exists, it would imply that flat space is indeed stable under RF.

In general a minimal surface can have any topology. 
Special interest attaches to the case where $S$ is an outermost 
horizon. In this case one can physically identify $S$ as the boundary of a 
black hole region. 
Outermost horizons
have the property that the surface has minimum area (and not just 
stationary area). This immediately implies \cite{gibbons} (when curvature 
is positive) that  the outermost horizon has spherical topology. Our 
general result shows that  
outermost  horizons {\it always} shrink under the Ricci flow. This 
result is diffeomorphism invariant.  If one identifies the area of an
outermost horizon as black hole entropy, we arrive at the conclusion that 
entropy is monotonically {\it decreasing} along the RF. A similar conclusion
has been reached by \cite{soludukhin} in a slightly different context:
entanglement entropy of a two dimensional black hole 
is monotonically decreasing along the RF.

If one uses the Ricci flow to model the approach of a system to thermal
equilibrium, one would expect that the entropy increases along the flow.
However, we have seen that outermost horizons have spherical topology and
their area {\it decreases} under the flow.  In fact, it has been observed 
\cite{black} that while the Ricci flow does have something in common
(memory loss) with approach to equilibrium, the analogy is not perfect.
A slight modification \cite{black} of the Ricci flow is necessary in order 
for the entropy to be identified with black hole entropy. 
It may be that such a modification of the 
RF is necessary for application to black hole physics.

This paper was concerned exclusively with the case of asymptotically
{\it flat} spaces. It would be interesting to generalise the treatment
to allow for asymptotically AdS spaces. This appears to be a straightforward
generalisation and some of the necessary changes are mentioned in 
\cite{gibbons,black}. 

\section*{Acknowledgements:} We thank Javed Ahmad for his collaboration
and Harish-Chandra Research Institute for their hospitality
in the early stages of this work. One of us (JS) thanks Harish Seshadri
for discussions on Ricci flows.


\begin{thebibliography}{99}
\section*{References}
\bibitem{friedan} Friedan D 1980 {\it Phys. Rev. Lett.} {\bf 45} 1057
\bibitem {hamilton} Hamilton R S 1982 Three-manifolds with positive 
Ricci curvature {\it J. Differential Geom.} 17
\bibitem {perelman} Perelman G 2002 The entropy formula for the Ricci 
flow and its geometric applications {\it Preprint} math.DG/0211159
\bibitem {cao} Cao H-D and Chow B 1999 Recent developments on 
the Ricci flow {\it Bull. Amer. Math. Soc.} {\bf 36} 59-74 
\bibitem {eric1} Woolgar E ``Some Applications of Ricci Flow in Physics"
eprint no arXiv:0708.2144 
\bibitem {eric2} Oliynyk T, Suneeta V and Woolgar E 
"A Metric for Gradient RG Flow of the Worldsheet Sigma Model"  
eprint: arXiv:0705.0827 to be published ({\it Phys. Rev. D.)}
\bibitem {eric3} Oliynyk T and Woolgar E 
"Asymptotically Flat Ricci Flows" arXiv:math/0607438
\bibitem {eric4} Oliynyk T, Suneeta V and Woolgar E 2006 "A gradient 
flow for world sheet non linear sigma models" {\it Nucl. Phys.B.} 
{\bf 739} 441-458
\bibitem {eric5} Oliynyk T, Suneeta V and Woolgar E 2005 
"Irreversibility of world sheet renormalisation group flow" {\it 
Phys.Lett. 
B.}
{\bf 610} 115-121
\bibitem {wiseman}
Headrick M and Wiseman T 2006 {\it Class.Quant.Grav.} {\bf 23} 6683-6708
\bibitem {soludukhin}
Soludukhin S 2007 {\it Phys.Lett. B.} {\bf 646} 268-274
\bibitem{black} Samuel J and Chowdhury S R 2007 Geometric 
Flows and Black Hole Entropy {\it Class. Quantum Grav.} {\bf 24} F1-F8
\bibitem {bekenstein}Bekenstein J D 1994 {\it Phys. Rev. D.} {\bf 9} 3292;
Bekenstein J D 1981 {\it Phys. Rev. D.} {\bf 23} 287
\bibitem{bousso}
Bousso R ``A Covariant Entropy Conjecture" 1999 {\it JHEP 9907} 004, 
arXiv:hep-th/9905177;
Bousso R ``The Holographic Principle" 2002 {\it Rev.Mod.Phys.} {\bf 74} 
825, arXiv:hep-th/0203101
\bibitem{unruh}
Unruh WG and Wald RM 1983 {\it Phys. Rev. D} {\bf 27}, 2271 - 2276
\bibitem{penrose} Penrose R 1973 {\it Ann. NY Acad. of Science} 224 
\bibitem{wald} Wald R M 1984 General Relativity 
\textit{University of Chicago Press}
\bibitem {bray} Bray H L 2003 {\it Preprint} math.DG/0304261
\bibitem {bray2} Bray H L 2001 {\it J. Diff. Geom.} {\bf 59} 177-267
\bibitem {huisken} Huisken G and Ilmanen T 2001 {\it J. Differential 
Geom.} {\bf 59} no.3
\bibitem{jang} Jang P S and Wald R M 1977 {\it J. Math. Phys.} {\bf 18} 41
\bibitem {geroch} Geroch R 1973 Energy Extraction {\it Ann. NY Acad. of 
Science} 224
\bibitem{topping} Topping P 2006 London Mathematical Society Lecture 
Notes Series {\it Cambridge University Press}
\bibitem{rfbook} Chow B and Knopf D 2004 The Ricci Flow: 
An Introduction {\it American Mathematical Society}
\bibitem {poisson} Poisson E 2004 {\it A Relativist's Toolkit} Cambridge 
Univ. Press
\bibitem{compact}
We use the word compact not in the mathematical sense, but in the physical 
sense as in ``A neutron star is a compact object''.
\bibitem{ivey}
Ivey TA ``The Ricci flow on radially symmetric $I\!\!R^3$" 1994 {\it Comm. 
Partial Differential eqns.} 1481
\bibitem{hayward} Hayward S A 1994 {\it Phys. Rev. D.} {\bf 49} 831  
\bibitem {schoen} Schoen R and Yau S-T 1979 Positivity 
of the Total Mass of a General Space-Time {\it Phys. Rev. Lett.} {\bf 43} 
20 
\bibitem {gibbons}
Gibbons G W 1999 {\it Class. Quant. Grav.} {\bf 16} 1677 
Gibbons G W 1996 Tunnelling with a negative cosmological constant {\it
Nucl. Phys. } {\bf B472} 683-708
Gibbons G W 1998 {\it Class. Quantum Grav.} {\bf 15} 2605-2612   
\end{thebibliography}
\end{document}